# Generalized theory of spin fluctuations in itinerant electron magnets: crucial role of spin anharmonicity

A. Solontsov[a,b]*

[a]Center for Fundamental and Applied Research, N.L. Dukhov Research Institute for Automatics, 22 Suschevskaya str., Moscow 127055, Russia,

[b]State Center for Condensed Matter Physics, 6/3 str. M. Zakharova, Moscow 115569, Russia

[*]e-mail: asolontsov@mail.ru

Dated: June 15, 2014

**Abstract**

The paper critically overviews the recent developments of the theory of spin fluctuations (SF) in itinerant electron magnetism with particular emphasis on spin-fluctuation coupling or spin anharmonicity. It is argued that the conventional self-consistent renormalized (SCR) theory of spin fluctuations is usually used aside of the range of its applicability actually defined by the constraint of weak spin anharmonicity based on the random phase approximation (RPA) arguments. An essential step in understanding SF in itinerant magnets beyond RPA-like arguments was made recently within the soft-mode theory of SF accounting for strong spin anharmonicity caused by zero-point SF. In the present paper we generalize it to apply for a wider range of temperatures and regimes of SF and show it to lead to qualitatively new results caused by zero-point effects.

## 1. Introduction.

Spin fluctuations (SF) in itinerant electron magnets are in the focus of condensed matter physics for more than half a century and play an important role in many classes of metallic systems including, e g., weak itinerant magnets [1], heavy fermion compounds [2], Invar alloys [3] and magnetoresistive manganites [4]. Recently they were also found in high-temperature superconductors [5] and newly discovered Fe-based superconductors [6] strongly suggesting their possible effect on mechanisms of unconventional superconductivity. However, up to now understanding of physics of SF in itinerant magnets and their role in superconductivity is not clear.

One of the most important problems of the SF theory in itinerant electron magnets is caused by strong coupling of SF (spin anharmonicity) induced by zero-point effects which cannot be treated within conventional perturbative schemes based on the random phase approximation (RPA) arguments (or Gaussian approximation). This problem is somewhat unique because SF are probably the only type of strongly coupled Bose excitations in the solid state physics. First

---

attempts to treat SF go back to 1960's when the theory of paramagnons, uncoupled overdamped Bose excitations in the electron-hole continua, was introduced based on the RPA (see Ref. 7). In 1970's Murata and Doniach [8] and Moriya and Kawabata [9] generalized the paramagnon theory in a self-consistent manner accounting for coupling of paramagnons. The formulated then a self-consistent renormalized (SCR) theory of SF established both microscopically and phenomenologically (see Refs.1,10) treated overdamped long wavelength SF basing on the RPA-like arguments. The initial version of the SCR theory where zero-point SF were neglected successfully explained many properties of weak itinerant magnets including the Curie-Weiss behaviour of the magnetic susceptibility [7,10]. The further version of the SCR theory partly incorporated zero-point SF and their temperature dependence [11,12]. However, the authors [11,12] used the same RPA arguments neglecting strong spin anharmonicity induced by zero-point effects. The improved version of the SCR theory was argued to lead to a simple renormalization of the parameters of the SCR theory treated within the phenomenological approach.

Another trend in the description of SF in itinerant magnets started in 1990's when it was realized that in the vicinity of magnetic instabilities SF soften similarly to softening of phonons near the structural transitions. However, unlike phonons soft SF give rise to giant amplitudes of zero-point SF [13] and to strong spin anharmonicity [14] which cannot be described within perturbative approaches based on the RPA arguments. To account for large zero-point SF amplitudes and strong spin anharmonicity a soft-mode (SM) theory of SF was formulated using both microscopic [15] and phenomenological [16] approaches. In the present paper we formulate a generalized theory of SF extending the SM theory [15-18] for a wider range of temperatures and regimes of SF. Zero-point SF are shown to lead to new qualitative effects including enhancement of low-temperature specific heat.

## 2. Model for spin fluctuations.

To account for various types of magnets with itinerant electrons instead of using microscopic models we start with the following phenomenological form for the inverse dynamical magnetic susceptibilities

$$\chi_\nu^{-1}(\mathbf{k},\omega) = \chi_\nu^{-1}(T) + c(\mathbf{k}) - i\frac{\omega}{\Gamma(\mathbf{k},\omega,T)} \qquad (1)$$

as a function of the wavevector $\mathbf{k}$, frequency $\omega$, temperature $T$ and polarization $\nu$ ($\nu = t$ marks the transverse and $\nu=l$ longitudinal ones), which was supported both theoretically and experimentally [1,7]. Here $\chi_\nu(T)$ are static susceptibilities, which coincide with the thermodynamic ones,

$$\chi_t^{-1} = \frac{1}{M}\frac{\partial F}{\partial M}, \quad \chi_l^{-1} = \frac{\partial^2 F}{\partial M^2}, \qquad (2)$$

$c(\mathbf{k})$ accounts for their spatial dispersion (we neglect its frequency and temperature dependencies which do not lead to new physical results), $F = F(T,M)$ is the free energy dependent on the order parameter $M$. Here $\Gamma(\mathbf{k},\omega,T)$ is the relaxation rate defining the nature of SF. At relatively low temperatures [18,19] SF relaxation is defined by the linear Landau mechanism in the electron-hole continua and is $\omega$- and $T$- independent. In this limit $\Gamma(\mathbf{k},\omega,T) \approx \Gamma_0(\mathbf{k})$ and SF have a conventional paramagnon nature. To account for the boundaries of electron-hole continua we introduce the wavevector $k_c$ and frequency $\omega_c$ cutoffs, where $k_c$ and $\omega_c$ are related to the electron-hole boundaries. For the Stoner continuum we also introduce a low frequency cutoff wavevector $k_0$ below which no transverse SF exist [10]. At elevated temperatures the SF relaxation mechanism is different from the linear Landau one and is defined by the various mode-mode scattering processes which dominate the magnetic relaxation [18,19] and may result in the frequency and temperature dependent relaxation rate $\Gamma(\mathbf{k},\omega,T)$ and lead to a number of novel phenomena [21,22]. Here for simplicity we shall not discuss the effects of non-linear magnetic relaxation assuming that the temperature is sufficiently low and set $\Gamma(\mathbf{k},\omega,T) \approx \Gamma_0(\mathbf{k}) \approx \Gamma_0 k$.

To find the free energy of itinerant electron magnets with strongly coupled SF models we adopt a physically transparent phenomenological approach based on the Ginsburg-Landau effective Hamiltonian

$$\hat{H}_{eff} = \frac{1}{2}\sum_{\mathbf{k}} \chi_0^{-1}(\mathbf{k})|\mathbf{M}(\mathbf{k})|^2 + \frac{\gamma_0}{4}\sum_{\mathbf{k}_1+\mathbf{k}_2+\mathbf{k}_3+\mathbf{k}_4=0}(\mathbf{M}(\mathbf{k}_1)\mathbf{M}(\mathbf{k}_2))(\mathbf{M}(\mathbf{k}_3)\mathbf{M}(\mathbf{k}_4)) \qquad (3)$$

which should be accompanied by the time-dependent equation

$$\frac{1}{\Gamma_0(\mathbf{k})}\frac{\partial \mathbf{M}(\mathbf{k})}{\partial t} = -\frac{\delta \hat{H}_{eff}}{\delta \mathbf{M}(-\mathbf{k})}. \qquad (4)$$

Here $\mathbf{M}(\mathbf{k},t) = \mathbf{M}\delta_{\mathbf{k},0} + \mathbf{m}(\mathbf{k},t)$ is the time dependent magnetic order parameter, $\mathbf{m}(\mathbf{k},t)$ accounts for SF, $\chi_0^{-1}(\mathbf{k}) = \chi_0^{-1} + c(\mathbf{k})$ and $\gamma_0$ define the static inhomogeneous paramagnetic susceptibility and mode-mode coupling constant not affected by SF. Here we treat (3) and (4) as a model describing SF and do not view them as expansions in powers of SF amplitudes which are not assumed to be small.

Then the free energy is given by

$$F(M,T) = F_0(T) + \frac{1}{2\chi_0}M^2 + \frac{\gamma_0}{4}M^4 + \Delta F, \qquad (5)$$

where $F_0(T)$ denotes the contribution independent on magnetization, terms with $M^2$ and $M^4$ are related to the Hartree-Fock approximation, and the SF contribution can be written using an integration of the equality $\partial \Delta F / \partial (\chi_0^{-1}) = M_L^2 / 2$,

$$\Delta F\{\chi_\nu(\mathbf{k},\omega)\} = \frac{1}{2}\sum_\nu \int d(\chi_0^{-1}) M_L^2 = \hbar \sum_\nu \sum_{\mathbf{k},\omega} \int d(\chi_0^{-1}) \operatorname{Im} \chi_\nu(\mathbf{k},\omega)[N_\omega + 1/2]. \qquad (6)$$

The squared local magnetic moment (averaged amplitudes of SF) $M_L^2 = \langle \mathbf{m}^2 \rangle$ is given by the fluctuation-dissipation theorem

$$M_L^2 = \sum_{\mathbf{k},\omega}(M_L^2)_{\mathbf{k},\omega} = 4\hbar \sum_\nu \sum_{\mathbf{k},\omega} \operatorname{Im}\chi_\nu(\mathbf{k},\omega)\left(N_\omega + \frac{1}{2}\right) \equiv \left(M_L^2\right)_{Z.P.} + \left(M_L^2\right)_T, \qquad (7)$$

where $\sum_{\mathbf{k},\omega} = \sum_\mathbf{k} \int_0^\infty (d\omega/2\pi)$, the factors $N_\omega = [\exp(\hbar\omega/k_B T) - 1]^{-1}$ and $1/2$ are related to thermal and zero-point SF, respectively, which provides a natural separation of $M_L^2$ into the zero-point $(M_L^2)_{Z.P.}$ and thermal $(M_L^2)_T$ contributions.

Integro-differencial equations (1),(2),(5), and (6) are the basic equations of the theory of SF that should be solved self-consistently. As it follows from Eq.( 6) the key parameter defining the solution of these equations is the derivative

$$\zeta_\nu = \frac{\partial(\chi_\nu^{-1})}{(\chi_0^{-1})} \qquad (8)$$

which is the measure of the effects of SF on the magnetic susceptibilities.

**3. Limitations of the self-consistent renormalized theory of spin fluctuations.**

In all versions of the SCR theory this parameter is set to unity,

$$\zeta_\nu = 1, \qquad (9)$$

which allows to present the SF contribution to the free energy (6) in the RPA-like form

$$\Delta F = 2\sum_\nu \sum_{\mathbf{k},\omega} F_{osc}(\omega) \frac{\omega_{SF}^{(\nu)}(\mathbf{k})}{\left[\omega_{sf}^{(\nu)}(\mathbf{k})\right]^2 + \omega^2} \equiv \sum_\nu \Delta F_{RPA}\{\chi_\nu(\mathbf{k},\omega)\}, \qquad (10)$$

where $F_{osc}(\omega) = k_B T \ln\left[1 - \exp(-\hbar\omega/k_B T)\right] + \hbar\omega/2$ is the free energy of a harmonic oscillator and

$$\omega_{SF}^{(\nu)}(\mathbf{k}) = \Gamma_0(\mathbf{k})[\chi_\nu^{-1} + c(\mathbf{k})]$$

(11)

are the relaxation frequencies of SF. The main result of the SCR theory is the shift of the magnetic susceptibilities [7] $\approx (5/3)\gamma_0 M_L^2$ resulting in the new derivation of the Curie temperature, Curie-Weiss law, etc. According to (8) and (9) this implies the main constraint for all versions of the SCR theory based on the RPA-like expression (10) for the SF free energy. Namely, the dimensionless spin anharmonicity parameter [14]

$$g_{SF} = \frac{\gamma_0}{3} \sum_\nu \left| \frac{\partial M_L^2}{\partial (\chi_\nu^{-1})} \right| = \frac{4}{3} \gamma_0 \hbar \sum_\nu \sum_{\mathbf{k},\omega} \operatorname{Im} \chi_\nu^2(\mathbf{k},\omega) \left( N_\omega + \frac{1}{2} \right) \tag{12}$$

must be small compared to unity,

$$g_{SF} = g_T + g_{Z.P.} \ll 1, \tag{13}$$

where according to the factor $N_\omega + 1/2$ in (12) we split it into the thermal $g_T$ and zero-point $g_{Z.P.}$ contributions, similar to the local moment (7).

As it follows from (12), the limit $g_{SF} = 0$ corresponds to the uncoupled Gaussian SF related to the paramagnon theories of 1960's. The initial version of the SCR theory [7,10] accounting for thermal SF and neglecting zero-point effects may be regarded as a first-order approximation in $g_{SF}$ which is valid only in the weak spin anharmonicity limit. The advanced version of the SCR theory partly accounting for zero-point SF within the RPA form for the free energy (10) will be discussed below.

**4. Soft-mode theory of spin fluctuations.**

However, the weak coupling constraint is hardly applicable to real itinerant magnets where zero-point SF were shown to have giant amplitudes both experimentally [13] and theoretically [16], which inevitably leads to strong spin anharmonicity breaking down the inequality (13) being the basis of the SCR theory. Assuming the linear Landau relaxation mechanism for SF one can easily estimate the squared local moment $M_{L0}^2 = [M_L^2(\chi_\nu^{-1} = 0)]_{Z.P.}$ and spin anharmonicity $g_0 = g_{Z.P.}(\chi_\nu^{-1} = 0)$ caused by zero-point SF [16]

$$M_{L0}^2 \sim \mu_B^2 N_e^2, \quad g_0 \sim \frac{M_{L0}^2}{\mu_B^2 N_e^2} \sim 1, \tag{14}$$

where $N_e$ is the electron density. Thus we arrive to an important conclusion: the conventional SCR theory of SF based on the RPA-like approximations for the free energy (10) does not account for the effects of strong spin anharmonicity caused by zero-point SF, which are intrinsic for itinerant electron magnets.

The way to take into account of strong anharmonicity of SF and to go beyond the RPA arguments was suggested within the SM theory of SF [14-18]. It proposed an outcome to solve self-consistently the equations (1), (2), (5) and (6) without the assumption (9), i.e. without using a perturbation approach. The main idea of the soft-mode theory is to consider the SM regime of SF, when the longitudinal susceptibility is less than the average spatial dispersion of thermal SF, $\chi_v^{-1} \ll c_T$, where [17] $c_T = c_c (T/T_{SF})^{2/3}$ if $T \leq T_{SF}$ and $c_T = c_c$ if $T > T_{SF}$. Here $c_c = c(k_c^2)$ is the measure of the spatial dispersion of SF, and $T_{SF} \approx \hbar \Gamma_0 c_c / k_B$ is the characteristic SF temperature which separates the temperature scale into the quantum ($T < T_{SF}$) and classical ($T > T_{SF}$) regions.

Within the SM theory self-consistent solution of equations (1), (2), (5), and (6) based on the expansions in powers of $\chi_v^{-1}$ gives the free energy of SF in the following form

$$\Delta F = \sum_v \frac{1}{\zeta_v} \Delta F_{RPA}\{\chi_v(\mathbf{k},\omega)\}, \quad (15)$$

which is enhanced with respect to the RPA expression (10) by the factors $1/\zeta_v > 1$. This allows to present the whole free energy (5) in the Landau form with the renormalized coefficients $\chi_0 \to \chi$, $\gamma_0 \to \gamma$, and $\zeta_v = 1 - 5g > 0$, where

$$\chi^{-1} = (1-5g)\chi_0^{-1} + \frac{5}{3}\gamma(M_{L0}^2 + M_T^2) < 0, \quad (16)$$

$$\frac{\gamma}{\gamma_0} = \frac{g}{g_0} = \frac{1-5g}{1+6g} \quad (17)$$

and $M_T^2 = [M_L^2(\chi_v^{-1}=0)]_T \sim M_{L0}^2 (T/T_{SF})^{4/3}$.

Here it is necessary to comment on the advanced version of the SCP theory which partly accounted for the zero-point SF basing on the RPA-like expression (10) for the free energy of SF [11,12]. As a result of their approach the authors [11,12] got the expression for the inverse paramagnetic susceptibility $\chi^{-1}$ which follows from (16) after setting $g = 0$ and changing $\gamma$ for a value essentially different from given by (17).

The main results of the SM theory were summarized in the reviews [17,18,22] and may be briefly formulated as follows. The SM theory:

i) clearly shows the limitations of the SCR theory which is valid in the weak anharmonicity limit $g_{SF} \ll 1$ within the SM regime (15) of SF;

ii) finds novel equation of states and criterion of magnetic instability with account of zero-point effects and strong spin anharmonicity;

iii) importantly, it founds fundamental limitations of the half-metallic ferromagnetic ground state due to quantum zero-point effects.

## 5. Generalization of the soft-mode theory.

Here we generalize the SM theory to discuss low-temperature anomalies of the specific heat caused by the effects of strong spin anharmonicity induced by zero-point SF in the Fermi-liquid regime of SF [17,18] which is out of the scope of the SM theory. Using the expressions (10) and (15) for the SF contribution to the free energy we get the SF specific heat $C_{SF} = -(1/T)\partial^2(\Delta F)/\partial M^2$ in the low-temperature limit

$$C_{SF} = \frac{k_B^2 T k_c^2}{12\pi\hbar\omega_{SF}(1-5g)}[\ln(1+\chi_l c_c) + 2\ln(\frac{k_c}{k_0})^2] = \frac{(C_{SF})_{RPA}}{1-5g}. \quad (18)$$

Here we take into account that in the ordered state of isotropic magnets $\chi_t^{-1} = 0$ and neglect the temperature dependence of the magnetic susceptibility $\chi_l$. As it is seen from (18), spin anharmonicity enhances the low-temperature specific heat by a factor $(1-5g)^{-1} > 1$ compared with the RPA result [20] $(C_{SF})_{RPA}$. The same result holds for the mass enhancement $m_{SF}/m = (m_{SF})_{RPA}/(1-5g)m$ caused by SF (where $m$ is the mass of a band electron inferred, e.g., from DFT calculations and $(m_{SF})_{RPA} \sim (C_{SF})_{RPA}$ is the RPA contribution to the electron mass from SF). This result should be taken into account while comparing calculated band structures of magnetic metals with experimentally observed data.

To conclude, we overview the existing spin fluctuation theories of itinerant electron magnets and generalize them to show that the low-temperature specific heat may be essentially affected by spin anharmonicity caused by zero-point effects.

## Acknowledgements

This work was supported by the State Atomic Energy Corporation of Russia "ROSATOM".